\documentclass[preprint,showpacs,preprintnumbers,amsmath,amssymb]{revtex4}
\usepackage{psfig}

\begin{document}
\title{Mean number of visits to sites 
in Levy flights }
\author      {M.Ferraro}
\affiliation    {Dipartimento  di Fisica Sperimentale
and CNISM,  \\
 via P.Giuria 1, I-10125 Turin,Italy }
\email {E-mail:ferraro@ph.unito.it}

\author     {L. Zaninetti}
\homepage   {http://www.ph.unito.it/~zaninett/index.html}    
\affiliation    {Dipartimento  di Fisica Generale, 
 via P.Giuria 1,\\ I-10125 Turin,Italy }
\email{zaninetti@ph.unito.it}
\date {\today}

\begin{abstract}

Formulas are derived to compute 
the mean number of times a site has been 
visited during symmetric Levy flights. 
Unrestricted Levy 
flights are considered first, for lattices of any dimension:
conditions for the existence of  finite 
asymptotic maps  of the visits over the lattice are analysed
and a connection is made with the transience of the flight.
In particular it is shown that flights 
on lattices of dimension greater than one are always transient.
For an interval with absorbing boundaries 
the mean number of visits reaches stationary values, which are computed 
by  means of  numerical and analytical methods; comparisons  
with  Monte Carlo  simulations  are also presented.

\end{abstract}

\maketitle  

\section{Introduction}

Levy flights  are a model of diffusion
in which the probability
of a $|z|$-length jump is ``broad'',  in that, asymptotically,
$p(z) \sim |z|^{-1-\alpha}$,  $0<\alpha <2 $.
In this case the 
sum $x_k=\sum_{i}^{k} z_i$ is distributed according 
to a Levy distribution,
 whereas for 
 $\alpha \geq 2$
normal 
diffusion takes place \cite{bib:gnko}, \cite{bib:boge}.
Interesting problems arise  in the theory
of Levy flights  
 when considering the statistics of
the visits to the sites, such for instance the number of 
different sites visited during a flight   \cite{bib:giwe}, \cite{bib:boha};
in this paper we consider a different, but related, problem,
namely the number of times a site  
visited by 
a random flyer.

Suppose that a random walk   takes place on a $d$-dimensional 
lattice 
$\mathcal L$, let ${\bf s}$ be a site of $\mathcal L$ 
and let 
$P^{(d)}_k({\bf s})$ be the probability that after $k$ steps the walker 
is at ${\bf s}$.
The mean value of visits to the site  ${\bf s}$ after $n$ 
steps is \cite{bib:feza}
\begin{equation}
M^{(d)}_n({\bf s})=\sum_{k=0}^{n}P^{(d)}_k({\bf s});
\label{eq:sola}
\end{equation}

since derivation of Eq. (\ref{eq:sola}) does not depend on the specific 
form of the 
walk \cite{bib:feza},  it holds  also for Levy flights. 
In the following it will be assumed 
$M_0({\bf s})=P_0({\bf s})=\delta_{{\bf s},0}$;
the asymptotic value of $M^{(d)}_n$, denoted by ${\mathcal M}^{(d)}$, 
is  defined as 
${\mathcal M}^{(d)}=\lim_{n \rightarrow \infty}M^{(d)}_n$. 
It is known  \cite{bib:fel} that a random walk is 
transient   if and only if $\sum_{k=0}^{\infty }P^{(d)}_k({\bf s}) < \infty $;
in other words  the existence of finite 
${\mathcal M}^{(d)}$ implies that the walk is transient.

Levy flights have a wide range of applications
(see for instance \cite{bib:ch} and references therein)
and, 
in particular, 
analysis of the number of times a site is visited 
can be relevant in those processes, such as random searches, 
 in which
 it is important 
 not just to determine what sites have been visited but 
how often they have been visited;  
examples of possible applications range from animal foraging
\cite{bib:visw} to  exploration of visual space 
\cite{bib:bocfer}.
Moreover  
$M_n^{(d)}$ can be given the following 
interpretation,  useful for possible applications:
assume that particles undergoing a Levy flight are continuously generated
at the origin, then, at time $n$,
 $M_n^{(d)}({\bf s}) \propto C_n^{(d)}({\bf s}) $, where 
$C_n^{(d)}({\bf s})$ is the mean number of particles   at site
${\bf s} \neq  0 $ 
\cite{bib:fzpha}. This property of $M^{(d)}_n$ has been used, 
in a model based on electrons Brownian motion, to simulate
 distributions of  emissivity of 
supernova remnants  \cite{bib:fzpha}.

\section{Infinite lattices}

Consider first one-dimensional, infinite 
lattices;  the probability of occupancy 
of site $x$ after $k$ steps is \cite{bib:hsm} 
\begin{equation}
\label{eq:conv}
P^{(1)}_{k+1}(x)=\sum_{x=-\infty}^\infty p(x-y)P^{(1)}_{k}(y),
\end{equation}

where $p(y)$ is  the probability of 
having a displacement of $y$ sites.
In case of symmetric Levy flights    
the canonical representation of $p$ and $P^{(1)}_k$
are \cite{bib:gnko}, \cite{bib:boge}

\begin{equation}
\label{eq:trans}
p(y)=\frac{1}{\pi}\int_0^\infty \cos qy \exp(-cq^\alpha) dq, 
\end{equation}

\begin{equation}
\label{eq:pro1}
P^{(1)}_k(x)=\frac{1}{\pi}\int_0^\infty \cos qx \exp(-ck q^\alpha) dq,
\end{equation}
 
 where $0 < \alpha < 2$ and  $c$ is a real number,  
which in the following  will be set equal to 1 
for simplicity
\cite{bib:boge}; a scaling relation holds 
between $P^{(1)}_k$ and $p$, namely  
$P^{(1)}_k(x)=k^{-1/\alpha}p(xk^{-1/\alpha})$.    
If $\alpha=2$ Eqs. (\ref{eq:trans}), (\ref{eq:pro1}) yield
 the  Gaussian distribution  \cite{bib:gnko},  \cite{bib:fel},
whereas, if $\alpha > 2$, $P^{1}_k$ 
fails to be a proper distribution not concentrated at a point 
\cite{bib:fel}; therefore  representations (\ref{eq:trans}),
(\ref{eq:pro1}) are valid only in the interval $0 < \alpha \leq 2$.
More recently it has been shown that    
the analytic forms  of $p$ and $P^{(1)}_k$
 can be  given through a Fox function
\cite{bib:mekl}.

Application of  (\ref{eq:sola}) 
and of  the  scaling relation leads to
$M^{(1)}_n(x)=\delta_{x,0}+\sum_{k=1}^n  k^{-\frac{1}{\alpha}} 
p\left(\frac{x}{k^{1/\alpha}}\right )$,
and in  particular, recalling that 
$p(0)=(\pi \alpha)^{-1}\Gamma(1/\alpha)$ \cite{bib:boge}, 

\begin{equation}
M^{(1)}_n(0)=1+\frac{\Gamma(1/\alpha)}{\pi \alpha}
\sum_{k=1}^n  k^{-\frac{1}{\alpha}},
\label{eq:zeta}
\end{equation}
  
with $\sum_{k=1}^n  k^{-1/\alpha}$ converging  to a finite value  for 
$n \rightarrow \infty$ if and only if $\alpha < 1$ \cite{bib:abst}; 
 in this case 
\begin{equation}
{\mathcal M}^{(1)}(0)=1+\frac{\Gamma(1/\alpha)}{\pi \alpha}\zeta(1/\alpha),
\label{eq:bizeta}
\end{equation}

where $\zeta$ is the well known Riemann zeta function \cite{bib:abst}.
Thus  Eqs. (\ref{eq:zeta}) and (\ref{eq:bizeta}),
show that the visit to site $x=0$ 
is a transient state if 
and only if $\alpha <1$.    

The trend of $M^{(1)}_n(0)$ 
as $n$ increases
can be computed 
by making use of 
the formulas related to the zeta function \cite{bib:abst};  
for $\alpha < 1 $ the result is

\begin{eqnarray}
&M&^{(1)}_n(0)=1+\frac{\Gamma(1/\alpha)}{\pi \alpha} \nonumber \\
&\times& \left (\zeta(1/\alpha)- \frac{\alpha}{1-\alpha} 
n^{\frac{\alpha-1}{\alpha}} + \frac{1}{\alpha} 
\int_n^\infty \frac{z-[z]}{z^{\frac{1}{\alpha}+1}}dz \right ), 
\label{eq:intor}
\end{eqnarray}

where $[z]$ is the integer part of $z$.
Application of standard summation formulas  \cite{bib:olv} shows that, 
if  $\alpha =1$, $M^{(1)}_n(0)$  grows logarithmically, whereas,
if $ 1 < \alpha <2 $, 

\begin{equation} 
M^{(1)}_n(0) \sim \frac{\Gamma(1/\alpha)}{\pi (\alpha-1)}
n^{\frac{\alpha-1}{\alpha}},
\label{eq:secfor}
\end{equation}

as  $n \rightarrow \infty$; 
finally in case of classical random walk ($\alpha \geq 2$), 
$M_n(0) = O(n^{1/2})$ \cite{bib:fzpha}.
Since flights are symmetric and  start from $0$, 
$P_k$ is, for every $k$,  an even function with a maximum in $0$ 
\cite{bib:mekl} and hence $M^{(1)}_n(0) > M^{(1)}_n(x)$,
for every $n$ and for every $x \neq 0$; 
therefore, if $\alpha < 1$, ${\mathcal M}^{(1)}(x) < \infty$.
A series expansion of Eq. (\ref{eq:pro1})
shows  that 
\begin{eqnarray}
\label{eq:ratio}
&&M^{(1)}_n(x)=M^{(1)}_n(0)- 1 \nonumber \\
&&+ \sum_{l=1}^{\infty}(-1^l)
\frac{\Gamma\left (\frac{2l+1}{\alpha} \right )}{\pi\alpha}
\frac{x^{2l}}{(2l)!}\sum_{k=1}^n k^{-\frac{2l+1}{\alpha}};
\end{eqnarray}

\noindent now  for every $0 <\alpha <2$ and 
every  $l$, $2l+1/\alpha >1$, and  $\lim_{n \rightarrow \infty}
 \sum_{k=1}^n k^{-\frac{2l+1}{\alpha}}=\zeta((2l+1)/\alpha)$ is finite.
Then
$M^{(1)}_n(x)= O(M^{(1)}_n(0))$, that is the last term 
on the RHS of (\ref{eq:ratio})
just takes into account the delay with which
 the flyer  reaches site $x$;
in particular, if  $\alpha >1$, $M^{(1)}_n(0)$ diverges and  
${M^{(1)}_n(x)} \sim {M^{(1)}_n(0)}$.
In conclusion,  a one-dimensional 
 flight is transient 
if and only if $\alpha < 1$, 
 a  result which  has been obtained in a somehow 
more complex way in \cite{bib:giwe}.

Consider now a $d$-dimensional lattice, with $d \geq 2$, and 
assume the  probabilities along the 
different coordinates to be independent; then  Eq. (\ref{eq:zeta} ) becomes  
$ M^{(d)}_n(0)=1 + \left (\frac{\Gamma(1/\alpha)}{\pi \alpha} \right )^d 
\sum_{k=1}^n k^{-\frac{d}{\alpha}}$. Note that, for  
$0 < \alpha < 2$ and $d \geq 2$, the condition   $d/\alpha >1$ holds and hence
 ${\mathcal M}^{(d)}(0)=1+\left(\frac{\Gamma(1/\alpha)}{\pi \alpha}\right)^d
\zeta(d/\alpha)$ is finite; 
$M^{(d)}_n(0)$ as a function of $n$ can be computed by using a method similar  
the one-dimensional case, and the result is that 
 the trend is given by 
$F(n)= \left (\frac{\Gamma(1/\alpha)}{\pi \alpha} \right )^d 
\left(n^{\frac{\alpha-d}{\alpha}}-1 \right) +O(n^{-\frac{d}{\alpha}})$.
Finally it should be observed that the results 
for $M^{(1)}_n(x)$,  $x \neq 0$, obtained above,  
can be extended in a straightforward  
way to multidimensional lattices.
Thus  Levy flights on lattices of  
dimensions higher than one are always 
transient; if $\alpha \geq 2$,  
$M^{(2)}_n(0)=O(\log (n))$ and,  if $d >2$, $M^{(d)}_n(0)$ 
converges to a finite value  
\cite{bib:fzpha}, and the walk is transient \cite{bib:fel}.

Note that, when $\alpha=1$,  $M^{(1)}_n(0)$ has the same trend  of 
$M^{(2)}_n(0)$ in the Gaussian regime, an instance of  
Levy flights  increasing  
the effective 
dimension of the walk \cite{bib:hsm}.  

\section{Finite intervals with absorbing boundaries}

In case of flights on a bounded set
it is  obvious that 
for  reflecting boundaries $M^{(d)}_n$ diverges as $n$ increases,
 since asymptotically 
$P^{(d)}_k \approx 1/|{\mathcal L}|$, where $|{\mathcal L|}$ 
is the number of sites \cite{bib:fel},
whereas if  boundaries are absorbing  
${\mathcal M}^{(d)}$ exists; here we shall consider 
just the case of one-dimensional lattices 
with absorbing boundaries.
The   map 
$M_n^{(1)}$ 
 can be computed 
by means of numerical or analytical methods. 
In fact,  Eq.  (\ref{eq:conv}) can be seen as a recursive method to 
compute $P^{(1)}_k$
 and application of (\ref{eq:sola}) provides the result;
alternatively, one can use the diffusion approximation
to derive an analytical formula.
 Both methods have been used here and their results   
have been compared with ${\mathcal S}(x)$,
the ``experimental'' number  of  visits  
 generated by  a Monte Carlo simulation.

In a closed interval  $[-a, a]$ Eq. (\ref{eq:conv}) becomes
\begin{equation} 
\label{eq:lconv}
P^{(1)}_{k+1}(x)=\sum_{x=-a}^a p(x-y)P^{(1)}_{k}(y),
\end{equation}

here, for reason of simplicity, instead of (\ref{eq:trans}), we have used  
the transition probability, defined on integers $y$, 
\begin{equation}
\label{eq:genlev}
 p(y) =\frac{1}{Z}|y|^{-(\alpha+1)}  \qquad {\rm if} \quad
 y \neq 0,
\end{equation}
and $p(0)=0$, $Z$ being a normalising constant.
A similar form of $p$ has been used in a work on 
the average time spent by flights in a closed interval \cite{bib:hav}. 
In case of numerical calculations, obviously, the absolute length $|y|$
 of a step must be truncated to some finite value:
here $\max(|y|)=2a$, to allow flights to  encompass the whole interval, 
 and consequently  
$Z=\sum_{y= -2a}^{2a}|y|^{-(\alpha+1)}$, $y \neq 0$.
Equation (\ref{eq:genlev}) provides a valid transition probability  
for any $\alpha >0 $ and hence it can be used to model 
also classical Brownian motion;  for $\alpha \rightarrow \infty $
the process becomes the simple symmetric walk.
Note that by combining (\ref{eq:conv}) and (\ref{eq:sola})
a recursive formula for $M^{(1)}_n$ can be derived, namely
$M^{(1)}_{n+1}(x)=\sum_{x=-a}^a p(x-y)M^{(1)}_{n}(y)+\delta_{x,0}$;
however the separate use of  (\ref{eq:conv}) and (\ref{eq:sola}) 
is to be preferred,  in that it also yields values of the probability 
distribution and this is useful to check the correctness of the results. 

In the classical theory of random walk 
the diffusion approximation
allows to replace $P^{(1)}_k({x})$ with the pdf  $P^{(1)}(x,t)$, solution  
of the  
diffusion  equation  \cite{bib:weiss}; analogously for Levy flights a 
superdiffusion equation 
can be derived (see, among others, 
\cite{bib:mekl}
\cite{bib:hav}, \cite{bib:gitt}), whose solution is 
a series of eigenfunctions $f_k$ of 
the operator ${\mathcal D}_\alpha$ \cite{bib:hav}.
Setting $P^{(1)}(x,0)=\delta(x-0)$, the pdf is   
$P^{(1)}(x,t)=\sum f_m(0)f_m(x)\exp(\lambda_mt)$. 
Define, in analogy with the discrete case,
\begin{equation} 
M^{(1)}(x,t)=\int_0^t P^{(1)}(x,\tau)d\tau, 
\label{eq:contin}
\end{equation}

then 
$M^{(1)}(x,t)=\sum_{m=1}^\infty \lambda^{-1}_mf_m(0)f_m(x)
(\exp(\lambda_m t)-1)$ where  
$\lambda_k$ are the eigenvalues of ${\mathcal D}_\alpha$; 
obviously,  $\lambda_k <0$, for all $k$, and the asymptotic formula is
${\mathcal M}^{(1)}(x)=\sum_{m=1}^\infty |\lambda_m|^{-1}{f_m(0)}f_m(x)$.

In \cite{bib:gitt} a solution $P^{(1)}(x,t)$ of the superdiffusion 
equation has been presented that, for symmetric flights, is 
\begin{eqnarray}
\label{eq:pgitt}
P^{(1)}(x,t)&=& \frac{2}{L}\sum_{m=1}^{\infty}
\exp\left[-D_\alpha(\pi m/L)^\alpha t\right] \nonumber \\
&\times& 
\sin\left (\frac{m\pi(x+a)}{L} \right ) \sin\left (\frac{m\pi a}{L}\right),
\end{eqnarray}
here  $L=2a$ is  the length of the interval and  $D_\alpha$ 
the diffusion coefficient;
application of Eq. (\ref{eq:contin}) to (\ref{eq:pgitt}), with 
$t \rightarrow \infty$, provides   
an explicit form for ${\mathcal M}^{(1)}(x)$,  
 
\begin{equation}
{\mathcal M}^{(1)}(x)=
\frac{2}{L}\sum_{m=1}^{\infty}\frac{L^\alpha}{(m \pi)^\alpha D_\alpha}
  \sin\left (\frac{m\pi(x+a)}{L} \right ) \sin\left 
(\frac{m\pi a}{L} \right ).
\label{eq:final}
\end{equation}

Calculations of ${\mathcal M}^{(1)}(x)$ from 
Eq.(\ref{eq:final}) need the numerical value 
of   the diffusion coefficient $D_{\alpha}$, 
and it  can be derived  from the average time $T$ a flyer 
spends in the interval, 
related to $D_{\alpha}$ by the formula \cite{bib:gitt}
\begin{equation}
\label{eq:expdi}
T=\frac{4}{\pi D_{\alpha}}
\left (\frac{L}{\pi} \right)^{\alpha} \sum_{m=1}^{\infty}
\frac{(-1)^m}{(2m+1)^{\alpha+1}};
\end{equation}

since $T$ is defined as    
$T=\int_{-a}^{a}dx\int_0^{\infty}P(x,t)dt
=\int_{-a}^{a}{\mathcal M}^{(1)}(x) dx$
the approximation 
$T \approx \sum_{x=-a}^a {\mathcal S}(x)$
can be used to obtain  the numerical value of $D_\alpha$.
% figure  theo_simu_0_3   
   \begin{figure}
\psfig{file=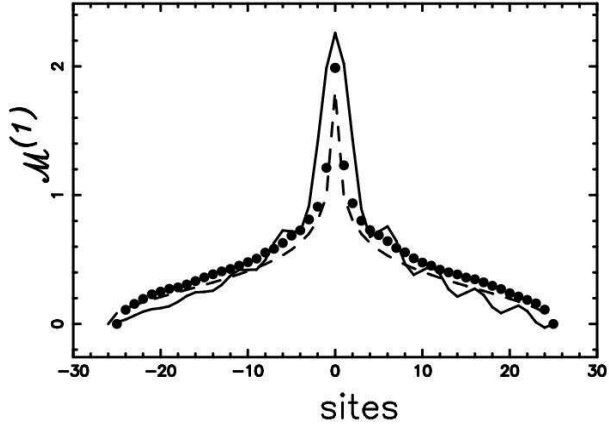,width=8 cm} 
\caption
{Graphs of ${\mathcal M}^{(1)}$, in case of absorbing boundaries:    
points denote the Monte Carlo simulation, 
the dashed line the numerical method 
via eqs.~\ref{eq:lconv} 
and~\ref{eq:sola}, and 
the full line the analytical solution (see~\ref{eq:final}).     
The  parameters are 
$\alpha=0.8$, $L=51$; 
The Monte Carlo simulation comprises  
$10000$ trials, in Eq. (\ref{eq:sola}) $n=2000$,
and the index $m$ in Eq. (\ref{eq:final})
ranges from $1$ to $20$ 
}
\label{theo_simu_0_3}%label 
    \end{figure}
% end figure  theo_simu_0_3 

% figure  theo_simu_2   
   \begin{figure}
\psfig{file=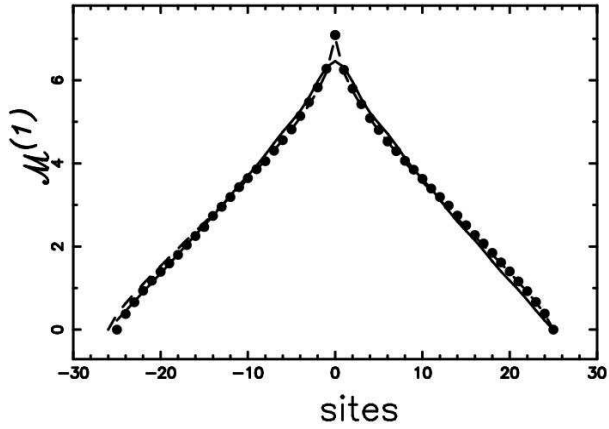,width=8 cm} 
\caption
{The same as Fig.~\ref{theo_simu_0_3}
but  $\alpha$=1.8
}
\label{theo_simu_2}%label 
    \end{figure}
% end figure  theo_simu_2 

Figures (\ref{theo_simu_0_3}) and (\ref{theo_simu_2}) 
show ${\mathcal M}^{(1)}$ for $\alpha=0.8$
and $\alpha=1.8$ respectively. 
It can be seen that the  graph of ${\mathcal M}$ tends
 to  a triangular shape as $\alpha$ increases; 
indeed for  simple  symmetric random walk, 
($\alpha=\infty$,  $p(x)=1/2\delta_{|x|,1}$) 
${\mathcal M}^{(1)}(x)=a-|x|$ \cite{bib:fzpha}.  
% figure  max_3
   \begin{figure}
\psfig{file=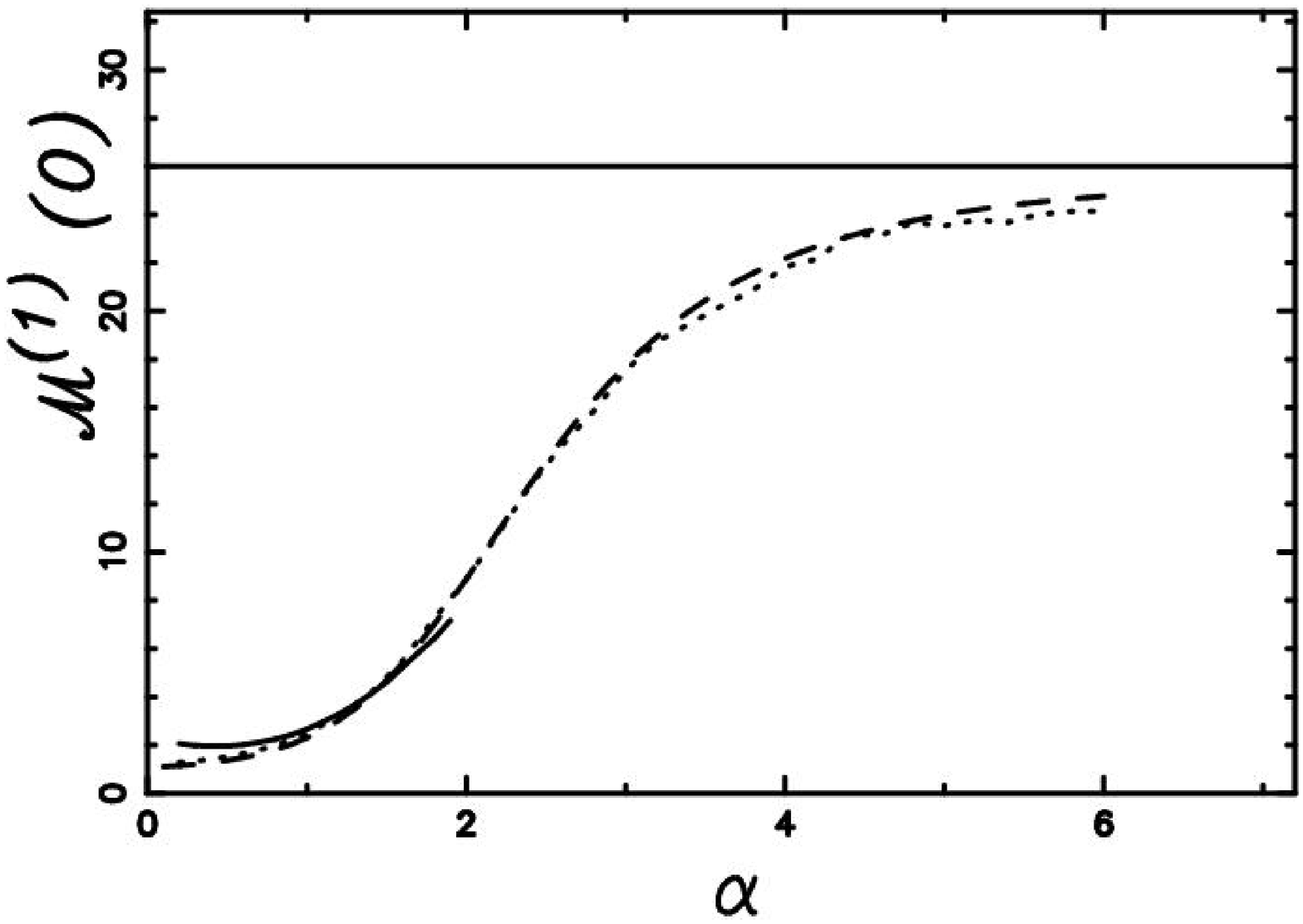,width=8 cm} 
\caption
{Graphs of ${\mathcal M}^{(1)}(0)$,  
as a function of  $\alpha$,
starting from $\alpha=0.2$,
in case of absorbing boundaries.   
Points represent the Monte Carlo simulation,  
the dashed line the numerical solution of~\ref{eq:lconv} 
and~\ref{eq:sola}, and the continuous line 
results from Eq.~(\ref{eq:final}),
with $\alpha <2$.  
The horizontal line is the result
for  $\alpha=\infty$  
}
\label{max_3}%label 
    \end{figure}
% end figure  max_3 

 Figure (\ref{max_3}) presents   
the graph of ${\mathcal M}^{(1)}(0)$ as a function of $\alpha$;
note that the inflection point of the curve occurs at $\alpha=2$, 
that is at the boundary between   
Levy flights and  classical random walks. In other words, 
${\mathcal M}^{(1)}(0)$
shows a ``phase transition'' 
from Levy flights, characterised by small  number of visits, 
to the Gaussian  regime where visits are  more frequent.
  
\section{Conclusion}
 The results of  this note clarify how the mean number of 
times a site is visited by a random flyer depends on the 
dimensionality of the lattice, the value of $\alpha$ and 
the boundary conditions. In particular, it has been shown that 
unrestricted Levy flights  are always transient, but for 
the unidimensional case with $\alpha \geq  1$; 
restricted flights are  transient if the boundaries are 
absorbing.
 In the last case computations show that 
  the direct numerical method agrees very closely with 
``experimental data'' generated by the Monte Carlo simulation, 
whereas the agreement is worse for Eq. (\ref{eq:final}),
especially when  $\alpha$ is small  
(see Figs.~\ref{theo_simu_0_3} and~\ref{max_3}); 
this is not surprising,  since Eq.~(\ref{eq:lconv}) 
deals directly with discrete variables, 
whereas Eq.~(\ref{eq:final}) results from the diffusion 
approximation.
On the other hand, obviously, Eq.~(\ref{eq:final})   
provides a more general, analytical formula for $\mathcal M^{(1)}$ 
and not just a set numerical values.\\

We thank  the two anonymous referees
for useful   advice and criticism.

\end{document}